\documentstyle[11pt,twoside]{article}

\begin{document}

\newcommand{\dd}{\,{\rm d}}
\newcommand{\ie}{{\it i.e.},\,}
\newcommand{\etal}{{\it et al.\ }}
\newcommand{\eg}{{\it e.g.},\,}
\newcommand{\cf}{{\it cf.\ }}
\newcommand{\vs}{{\it vs.\ }}
\newcommand{\zdot}{\makebox[0pt][l]{.}}
\newcommand{\up}[1]{\ifmmode^{\rm #1}\else$^{\rm #1}$\fi}
\newcommand{\dn}[1]{\ifmmode_{\rm #1}\else$_{\rm #1}$\fi}
\newcommand{\upd}{\up{d}}
\newcommand{\uph}{\up{h}}
\newcommand{\upm}{\up{m}}
\newcommand{\ups}{\up{s}}
\newcommand{\arcd}{\ifmmode^{\circ}\else$^{\circ}$\fi}
\newcommand{\arcm}{\ifmmode{'}\else$'$\fi}
\newcommand{\arcs}{\ifmmode{''}\else$''$\fi}
\newcommand{\MS}{{\rm M}\ifmmode_{\odot}\else$_{\odot}$\fi}
\newcommand{\RS}{{\rm R}\ifmmode_{\odot}\else$_{\odot}$\fi}
\newcommand{\LS}{{\rm L}\ifmmode_{\odot}\else$_{\odot}$\fi}

\newcommand{\Abstract}[1]{{\footnotesize\begin{center}ABSTRACT\end{center}
\vspace{1mm}\par#1\par}}

\newcommand{\TabCap}[2]{\begin{center}\parbox[t]{#1}{\begin{center}
  \small {\spaceskip 2pt plus 1pt minus 1pt T a b l e}
  \refstepcounter{table}\thetable \\[2mm]
  \footnotesize #2 \end{center}}\end{center}}

\newcommand{\TableSep}[2]{\begin{table}[p]\vspace{#1}
\TabCap{#2}\end{table}}

\newcommand{\TableFont}{\footnotesize}
\newcommand{\TableFontIt}{\ttit}
\newcommand{\SetTableFont}[1]{\renewcommand{\TableFont}{#1}}

\newcommand{\MakeTable}[4]{\begin{table}[htb]\TabCap{#2}{#3}
  \begin{center} \TableFont \begin{tabular}{#1} #4 
  \end{tabular}\end{center}\end{table}}

\newcommand{\MakeTableSep}[4]{\begin{table}[p]\TabCap{#2}{#3}
  \begin{center} \TableFont \begin{tabular}{#1} #4 
  \end{tabular}\end{center}\end{table}}

\newenvironment{references}%
{
\footnotesize \frenchspacing
\renewcommand{\thesection}{}
\renewcommand{\in}{{\rm in }}
\renewcommand{\AA}{Astron.\ Astrophys.}
\newcommand{\AAS}{Astron.~Astrophys.~Suppl.~Ser.}
\newcommand{\ApJ}{Astrophys.\ J.}
\newcommand{\ApJS}{Astrophys.\ J.~Suppl.~Ser.}
\newcommand{\ApJL}{Astrophys.\ J.~Letters}
\newcommand{\AJ}{Astron.\ J.}
\newcommand{\IBVS}{IBVS}
\newcommand{\PASP}{P.A.S.P.}
\newcommand{\Acta}{Acta Astron.}
\newcommand{\MNRAS}{MNRAS}
\renewcommand{\and}{{\rm and }}
\section{{\rm REFERENCES}}
\sloppy \hyphenpenalty10000
\begin{list}{}{\leftmargin1cm\listparindent-1cm
\itemindent\listparindent\parsep0pt\itemsep0pt}}%
{\end{list}\vspace{2mm}}

\def\TYLDA{~}
\newlength{\DW}
\settowidth{\DW}{0}
\newcommand{\dw}{\hspace{\DW}}

\newcommand{\refitem}[5]{\item[]{#1} #2%
\def\REFARG{#3}\ifx\REFARG\TYLDA\else, {\it#3}\fi
\def\REFARG{#4}\ifx\REFARG\TYLDA\else, {\bf#4}\fi
\def\REFARG{#5}\ifx\REFARG\TYLDA\else, {#5}\fi.}

\newcommand{\Section}[1]{\section{#1}}
\newcommand{\Subsection}[1]{\subsection{#1}}
\newcommand{\Acknow}[1]{\par\vspace{5mm}{\bf Acknowledgements.} #1}
\pagestyle{myheadings}

\newfont{\bb}{ptmbi at 12pt}

\def\thefootnote{\fnsymbol{footnote}}

\begin{center}
{\Large\bf The Optical Gravitational Lensing Experiment.\\
\vskip3pt
The Distance Scale:\\
\vskip3pt
Galactic Bulge -- LMC -- SMC.
\footnote{Based on  observations obtained with the 1.3~m Warsaw
telescope at the Las Campanas  Observatory of the Carnegie Institution
of Washington.}}
\vskip1cm
{\bf A.~~U~d~a~l~s~k~i$^1$}
\vskip5mm
{$^1$Warsaw University Observatory, Al.~Ujazdowskie~4, 00-478~Warszawa, Poland\\
e-mail: udalski@sirius.astrouw.edu.pl }
\end{center}
\vskip 10mm

\Abstract{We analyze the mean luminosity of three samples of field
RRab~Lyr  stars observed in the course of the OGLE microlensing
experiment: 73 stars  from the Galactic bulge and 110 and 128 stars from
selected fields in the LMC  and SMC, respectively. The fields are the
same as in the recent distance  determination to the Magellanic Clouds
with the red clump stars method by Udalski \etal (1998). We determine 
the relative distance scale $d_{GB}^{RR}:d_{LMC}^{RR}:d_{SMC}^{RR}$
equal to:  ($0.194\pm0.010):1.00:(1.30\pm0.08$). 

We calibrate our RR~Lyr distance scale with the recent calibration of
Gould and Popowski (1998) based on statistical parallaxes. We obtain
the  following distance moduli to the Galactic bulge, LMC and SMC: ${m-
M=14.53{\pm}0.15}$, ${m-M=18.09{\pm}0.16}$ and ${m-M=18.66\pm0.16}$~mag.

We use the RR~Lyr mean {\it V}-band luminosity at the Galactic bulge
metallicity as the  reference brightness and analyze the mean, {\it
I}-band luminosity of the red  clump stars in objects with different
ages and metallicities. We add to our analysis the metal poor Carina
dwarf galaxy which contains old RR~Lyr stars and intermediate age red
clump population. We find a weak dependence of the mean red clump
brightness on metallicity and we calibrate its zero point with the
nearby local red clump stars measured by Hipparcos: $M^{\rm
RC}_I=(0.09\pm0.03)\times{\rm [Fe/H]^{\rm RC}}-0.23\pm0.03$. Our revised
red clump distance moduli to the Galactic bulge, LMC and SMC are ${m-
M=14.53\pm0.06}$, ${m-M=18.13\pm0.07}$ and ${m-M=18.63\pm0.07}$~mag,
respectively. The distance modulus to the Carina galaxy is
${m-M=19.84\pm0.07}$~mag. Excellent  agreement of RR~Lyr and red clump
distances  which have independent absolute calibrations confirms the
short distance scale to the LMC.}

\Section{Introduction}
The distance scale to the Magellanic Clouds is one of the most important
problem of the modern astrophysics. In particular the distance to the
Large  Magellanic Cloud is a standard distance to which the
extragalactic distance  scale based on the Cepheid period-luminosity
(P--L) relation is tied.  Therefore any error in determination of the
LMC distance propagates  immediately to the extragalactic distances, to any
value which is scaled by  distance and also to the Hubble constant. Thus,
precise determination of the LMC  distance is of the greatest
importance. 

Unfortunately the distance to the LMC has for a long time been a subject
of  controversies. The most widely accepted "long" distance scale,
${m-M\approx  18.50}$~mag, was also adopted by the HST Extragalactic
Distance Scale team.  The "long" distance scale to the LMC is based
mostly on the Cepheid P--L relation  determinations (Madore and Freedman
1998) and some determinations from the  supernova SN1987A light echo
(Panagia \etal 1997). 

On the other hand observations of other standard candle -- RR~Lyr stars
--  seem to favor much shorter distance to the LMC,
${m-M\approx18.3}$~mag.  (Layden \etal 1996). Such a "short" scale also
seems to be preferred by  reanalysis of the supernova SN1987A echo
(Gould and Uza 1998). 

The distance to the Small Magellanic Cloud is poorly known because
observations of  this galaxy with modern techniques are rare. The widely
accepted SMC distance  modulus is ${m-M=18.94}$~mag from the Cepheid
P--L relation (Laney and Stobie  1994). 

Recently Udalski \etal (1998) employed a new technique of distance 
determination, proposed by Paczy{\'n}ski and Stanek (1998), to determine
distances to the Magellanic Clouds using the mean {\it I}-band
luminosity of the red clump stars as a standard candle. The red clump is
an  equivalent of the horizontal branch for intermediate population 
(age 2--10~Gyr) He-burning stars and is present in galaxies or clusters 
possessing such a population of stars. 

The main advantage of the red clump stars method is that it gives a
possibility of precise  determination of the mean red clump luminosity,
as the red clump stars are  very numerous. Moreover, the red clump stars
are also very numerous in the solar  neighborhood and therefore the
method can be calibrated very precisely with  nearby stars for which
parallaxes were measured with high precision by the Hipparcos satellite.
No other standard candle calibration is based on direct  measurements of
hundreds of individual stars with precision better than 10\%.  Thus the
red clump, single-step method seems to be potentially the most  precise
technique of distance determination used so far. 

Udalski \etal (1998) found ${m-M=18.08\pm0.03\pm0.12}$~mag and $m-
M=18.56\pm0.03\pm0.06$~mag for the LMC and SMC, respectively (errors are
statistical and systematic, mostly from interstellar extinction
uncertainty,  respectively). Stanek, Zaritsky and Harris (1998) fully
confirmed results of  Udalski \etal (1998). Using independent photometry
they obtained the distance  modulus to the LMC equal to
${m-M=18.07\pm0.03\pm0.09}$~mag reducing the  systematic error by using
accurate maps of interstellar extinction of their  fields. 

The most controversial assumption of the red clump method is the
constant  mean {\it I}-band brightness of the red clump stars.
Paczy{\'n}ski  and Stanek (1998), Udalski \etal (1998) and Stanek,
Zaritsky and Harris (1998)  discussed in detail the evidences in support
of such an assumption, at least  in the first approximation. On the
other hand Cole (1998) suggested large,  reaching ${\approx0.6}$~mag,
variations of the mean red clump magnitude  resulting from population
differences (age, metallicity) in different objects.  He attempted to
calibrate the mean absolute, {\it I}-band magnitude of the red  clump
stars based on evolutionary models of the red clump stars and formation 
history of the Magellanic Clouds. Girardi \etal (1998) analyzed
evolutionary models of the red clump stars and found the difference of
$\Delta M_I=-0.235$ mag between the absolute brightness of the red clump
in the LMC  and local Hipparcos measured sample. This theoretical
approach involves, however, many difficult to verify and often
controversial assumptions  like mass loss, helium content, stars
formation rate etc. 

In this and subsequent papers of this series we propose another, fully
empirical approach to the problem of verification  how much  population
effects can affect the mean {\it I}-band luminosity of the red  clump
stars. In this paper we present and analyze large samples of RR~Lyr
stars from the LMC and SMC obtained as a by-product of the second phase
of the Optical  Gravitational Lensing Experiment, OGLE-II, (Udalski,
Kubiak and Szyma{\'n}ski  1997). The samples come from the same fields
which were analyzed with the red  clump stars method. Together with the
large sample of RR~Lyr stars from the  Baade's Window region of the
Galactic bulge obtained during the first phase of  the OGLE experiment,
OGLE-I, we derive the relative RR~Lyr distance scale between the 
Galactic bulge, LMC and SMC and determine absolute distances with the
best present RR~Lyr calibration.

With two standard candles in objects of different ages and metallicities
(additionally we include in our analysis the Carina dwarf galaxy which
contains similar mixture of different age stellar populations), we may
treat the  mean brightness of RR~Lyr stars as a reference point and
check how the mean  luminosity of the red clump stars changes from
object to object. We find a  clear trend which allows us to recalibrate
the mean absolute {\it I}-band  magnitude of the red clump and determine
revised red clump distances.

\Section{Observational Data}
All observations of the Magellanic Cloud RR~Lyr stars were obtained with
the  1.3-m Warsaw telescope located at the Las Campanas Observatory
which is  operated by the Carnegie Institution of Washington during the
second phase of  the OGLE project (Udalski, Kubiak and Szyma{\'n}ski
1997). 

The SMC objects were extracted from the preliminary version of the SMC
RR~Lyr  stars catalog which will be released later this year. Stars from
the same  fields which were used for the red clump distance
determination were  extracted, namely SMC$\_$SC1, SMC$\_$SC2,
SMC$\_$SC10 and SMC$\_$SC11 (Udalski  \etal 1998). Each of these fields
cover ${\approx14\arcm\times56\arcm}$ on the  sky. Observations spanned
the period of Jun.~26, 1997 through Feb.~12, 1998.  About 100--115 {\it
I}-band observations were used for RR~Lyr search.  Additionally about
20--25 {\it V}-band observations were used for  determination of the
mean ${V-I}$ color of the RR~Lyr stars. 128 RR~Lyr of ab type  were
selected in the SMC fields. 

It is important to note that one of the RR~Lyr stars,
SMC$\_$SC11~101421, is  located only 87~arcsec from the center of the
rich cluster NGC~419.  It is possible that this is just a coincidence,
because the NGC~419 is a  relatively young cluster with the age of about
3~Gyr (Bica, Dottori and  Pastoriza 1987) and should not harbor stars as
old as RR~Lyr. Only the oldest  cluster in the SMC, NGC~121 contains a
few RR~Lyr stars (Olszewski,  Suntzeff and Mateo 1996). Nevertheless,
because of potential consequences on  our understanding of cluster
population in the SMC a possibility of cluster  membership of
SMC$\_$SC11~101421 should be carefully reinvestigated. 

The LMC RR~Lyr stars were extracted from preliminary databases of the
LMC  fields: LMC$\_$SC14, LMC$\_$SC15, LMC$\_$SC19 and LMC$\_$SC20.
These fields  were added to the main OGLE-II LMC fields at the beginning
of the 1997/98  season, and therefore the number of collected
observations is smaller  -- about 65 in the {\it I}-band and only 6 in
the {\it V}-band. Observations  presented in this paper cover the period
Oct.~5, 1997 through Apr.~22, 1998. To  avoid differential extinction
only 1/3 of each LMC field was searched --  similar as for the red clump
distance determination (see Table~1, Udalski  \etal 1998). In total 110
ab type RR~Lyr stars were found in the LMC  fields. 

All LMC and SMC fields were carefully calibrated and accuracy of the
zero  points of the OGLE photometry is about 0.01~mag for the SMC  and
0.01--0.02~mag for the LMC fields (Udalski \etal 1998). Figs.~1 and 2
present sample light curves of RRab Lyr stars from the SMC and LMC,
respectively, and Table~1 and 2  list basic parameters of all RR~Lyr
stars. 

The Galactic bulge RR~Lyr stars data come from the OGLE-I catalogs of
periodic  variable stars (Udalski \etal 1994, 1995a 1995b) which are
available in digital form from the OGLE archive: {\it
ftp://sirius.astrouw.edu.pl/ogle/ /var\_catalog} or {\it
http://www.astrouw.edu.pl/\~{}ftp/ogle}. Only RRab type stars from  fields
BW1-8 and BWC for which interstellar extinction could be derived from
Stanek (1996) extinction map, in total 73 objects, were selected.
Accuracy of  the zero points of the OGLE-I photometry in dense bulge
fields is about  0.03~mag (Udalski \etal 1993). Light curves of Galactic
bulge RRab stars can  be found in Udalski \etal (1994, 1995a 1995b).
Table~3 lists the most  important parameters of the Galactic bulge
RRab~Lyr stars.

Observations of the additional object analyzed in this paper -- the
Carina dwarf galaxy -- were obtained as a side-project of the OGLE-II
experiment. {\it VI}-band observations of this galaxy started on
Apr.~16, 1998 and are still being continued. So far about 20 {\it VI}
images of the field centered at ${\rm RA}_{2000}=6\uph41\upm34\ups$ and 
${\rm DEC}_{2000}=-50\arcd58\arcm00\arcs$ were collected. The data,
contrary to the Magellanic Clouds driftscan mode observations, are
obtained in the normal, still frame mode and cover about
${14\arcm\times14\arcm}$ on the sky. Typical exposure times are
600 and 900 sec for  {\it V} and {\it I}-band, respectively.
Observations were calibrated with the measurements of several standard
stars from Landolt (1992) list collected on four photometric nights.
Accuracy of the zero points of the absolute photometry is about
0.015~mag. Comparison of our photometry with previous studies (location
of the red clump stars - see Section~4) shows excellent agreement with
Hurley-Keller, Mateo and Nemec (1998) data for the {\it V}-band,
Smecker-Hane \etal (1994) {\it I}-band observations, and Mighell (1997)
HST WFPC2 {\it VI}-band data.

\Section{RR~Lyrae Distance Scale: Galactic Bulge -- LMC -- SMC} 

\Subsection{RR~Lyrae Data}

To determine the distance scale to the Galactic Bulge, LMC and SMC with
RR~Lyr  stars we calculated the mean magnitudes of RR~Lyr stars in our
three  samples. First, we derived the mean, {\it I}-band intensity
weighted magnitudes of each  star in the sample. We fitted the light
curve with high order polynomials and determined the mean intensity. We
checked the results by repeating the procedure with approximation of the
light curve with Fourier series. Results were consistent at the 0.01 mag
level.

Then we corrected the observed magnitudes for interstellar  extinction.
In the case of the Galactic bulge stars we used Stanek (1996) 
extinction maps with the zero point correction determined by Gould,
Popowski  and Terndrup (1998) and Alcock \etal (1998b). For the
Magellanic Cloud fields we  assumed the same extinction as for the red
clump stars:  ${A_I=0.33}$~mag for the LMC$\_$SC14 and LMC$\_$SC15
fields, ${A_I=0.39}$~mag for  the LMC$\_$SC19 and LMC$\_$SC20 fields and
${A_I=0.16}$~mag for the SMC fields (Udalski \etal 1998).  Good
agreement of the red clump distance to the LMC derived by Udalski \etal
(1998) with  Stanek, Zaritsky and Harris (1998) determination for other
fields confirms that these estimates of extinction are correct.

In the next step we prepared histograms of the mean magnitude of the
RR~Lyr stars from each sample  in 0.07~mag bins. In the case of the
Magellanic Clouds we decided to merge samples  of fields located in
west- and eastward parts of each Magellanic Cloud to increase statistics
(hereafter called LMC$\_$E, LMC$\_$W, SMC$\_$E and SMC$\_$W fields).
Analysis of the red clump (Udalski \etal 1998) indicates that in the 
merged fields the extinction is similar which fully justifies this step.
Next, we fitted a Gaussian to each  sample. The mean magnitude, $I_0^{\rm
RR}$, its  statistical  error, $\sigma_{I_0}^{\rm ST}$, systematic
error, $\sigma_{I_0}^{\rm SYS}$, dispersion of the Gaussian,
$\sigma_{\rm RR}$, and number of stars in each sample are given in
Table~4. We also  calculated the mean and median magnitude and standard
deviation of each sample  to compare with results of Gaussian modeling.
Results were almost indistinguishable,  only the standard deviation was
in some cases about ${0.01-0.02}$~mag larger  due to a few outliers.

\setcounter{table}{3}
\MakeTable{lccccc}{12.5cm}{The mean {\it I}-band magnitudes of RR~Lyr stars}
{
         & $N_*$ & $I_0^{\rm RR}$ & $\sigma_{I_0}^{\rm ST}$ & 
           $\sigma_{I_0}^{\rm SYS}$ & $\sigma_{\rm RR}$ \\
&&&&&\\
GB       &  73   & 14.77 & 0.019  & 0.04  & 0.25\\
LMC$\_$E &  43   & 18.45 & 0.020  & 0.05  & 0.19\\
LMC$\_$W &  67   & 18.38 & 0.014  & 0.05  & 0.14\\
SMC$\_$E &  59   & 18.92 & 0.026  & 0.05  & 0.21\\
SMC$\_$W &  69   & 18.93 & 0.014  & 0.05  & 0.18\\
}

Figs.~3 and 4 present histograms with a Gaussian function fitted to the
data. As  can be seen from Table~4 the $\sigma_{\rm RR}$ for each sample
is similar to  that of the red clump stars (Udalski \etal 1998,
Paczy{\'n}ski and Stanek  1998) confirming that RR~Lyr stars can be used
as a good standard candle.  $\sigma_{\rm RR}$ for the Magellanic Cloud
samples is in good agreement with that  of samples from other regions of
the SMC and LMC (Smith \etal 1992, Hazen  and Nemec 1992). Dispersion
larger than observed in the globular clusters  (typically ${\sigma_{\rm
RR}\approx0.1}$~mag) can be explained, at least in part, by depth 
effects for the field RR~Lyr stars from our samples. The mean magnitude,
$I_0^{\rm RR}$, is determined with larger  statistical error than that
of the red clump due to much smaller number of stars. However, our
RR~Lyr samples will be increased by order of magnitude  when precise
extinction maps are derived from the OGLE-II data. In this  paper we
limited ourselves to stars located only in fields for which the red 
clump stars distances were determined.

The mean {\it I}-band magnitudes of the RR~Lyr samples from different
lines-of-sight in the  SMC are in very good agreement. There is a few
hundredth of magnitude difference  for the two LMC locations. It is not
clear if it is real and therefore we  adopt the mean as the LMC
mean RR~Lyr magnitude. Adopted mean {\it I}-band magnitudes of  the
RR~Lyr stars for the Galactic bulge, LMC and SMC are listed in Table~5.

To determine the mean {\it V}-band magnitudes of RR~Lyr stars which are 
necessary for the distance scale determination we derived the mean {\it
V-I} colors of our samples. Individual color measurements were obtained
by subtracting the {\it I}-band magnitude calculated from the polynomial
fit to the {\it I}-band light curve at the {\it V}-band observation
phase from the observed {\it V}-band magnitude. Such color curves were
then averaged. In the case of the LMC RR~Lyr stars for which number of
{\it V}-band measurements is small, the mean color is somewhat less
accurate. However, our extensive tests showed that with a sample of
RR~Lyr as numerous as our LMC sample the mean color is practically
stable when more than four individual observations per star are
collected. Our color curves of the LMC fields have six observations and
therefore we estimate that the mean color error should not exceed
0.03~mag. Similar $(V-I)_0$  color of the LMC RRab~Lyr stars from the 
NGC~2210 region  ($(V-I)_0=0.48$~mag) was obtained by Reid and Freedman
(1994).

Mean colors of the LMC and SMC RR~Lyr stars were corrected for reddening
using the same values as for the red clump star distance determination:
$E(V-I)=0.10$ mag for the SMC fields, 0.22~mag for the LMC$\_$W and
0.26~mag for the LMC$\_$E fields (Udalski \etal 1998). Color of each
RR~Lyr star from the Galactic bulge was individually corrected for
reddening with Stanek (1996) extinction map before determination of the
mean $(V-I)_0$ color of the sample.

\MakeTable{lcccccccc}{12.5cm}{Mean {\it V, I} and {\it V-I} extinction
free magnitudes of RR~Lyr stars}
{
          & $I_0^{\rm RR}$ & $\sigma_{I_0}^{\rm ST}$ & $\sigma_{I_0}^{\rm SYS}$&

         $(V-I)_0^{\rm RR}$ & $\sigma_{(V-I)_0}^{\rm ST}$ &
         $V_0^{\rm RR}$ & $\sigma_{V_0}^{\rm ST}$ & $\sigma_{V_0}^{\rm SYS}$ \\
&&&&&&&&\\
GB        & 14.77 & 0.02 & 0.04 & 0.64 & 0.02 & 15.41 & 0.03 & 0.06 \\
LMC       & 18.41 & 0.02 & 0.05 & 0.45 & 0.03 & 18.86 & 0.04 & 0.08 \\
SMC       & 18.93 & 0.02 & 0.05 & 0.48 & 0.02 & 19.41 & 0.03 & 0.08 \\
Carina DG & 19.96 & 0.03 & 0.04 & 0.53 & 0.03 & 20.49 & 0.04 & 0.06 \\
}

Table~5 lists the mean ${\it (V-I)}_0$ color of each of our samples with
its statistical error as well as the mean ${\it V}_0$ magnitudes of
RR~Lyr stars with statistical and systematic (extinction uncertainty)
errors.

In Table~5 we provide also the corresponding data for the Carina dwarf
galaxy, determined from our observations. These data will be used below
for analysis of the red clump stars. As the number of collected
observations is too small to perform a reasonable search for RR~Lyr
stars in our Carina field, we decided to limit ourselves to  RR~Lyr
stars detected in our field by Saha, Monet and Seitzer (1986). We
identified 15 RRab Lyr stars and determined their mean {\it V} and {\it
I}-band magnitudes by averaging their intensities. Then we corrected the
mean magnitudes for interstellar extinction assuming the reddening to
the Carina dwarf galaxy equal to $E(V-I)=0.08\pm0.02$ (Mighell 1997) and
standard relations between extinction in the {\it V} and {\it I}-bands
and $E(V-I)$.

The important question when comparing objects from different locations
is  completeness of the sample, especially when statistics are not large
and  any bias due to incompleteness may lead to gross error. In the case
of the  Galactic bulge, which RR~Lyr stars are much brighter and light
curves  determined with very good precision the situation is relatively
simple. Tests  of completeness of the OGLE-I catalogs performed on the
overlapping regions of  a few fields showed that completeness for RR~Lyr
stars is  practically 100\% (Udalski \etal 1995b). We excluded from our
Galactic bulge  sample of RRab stars seven objects for which extinction
could not be  determined from the Stanek (1996) map.

In the case of the Magellanic Clouds the chance of overlooking some
faint 19-th  magnitude objects is larger because of much more noisy
light curves. However,  the RRab~Lyr stars which are large amplitude
(${>0.4}$~mag) variable stars,  should be rather easily detectable down
to ${I\approx20}$~mag. Although tests  of the completeness of the OGLE-II
variable stars catalogs have not been  performed yet, we might estimate
that the completeness for RRab type stars is  also high and close to
100\%. 

Smith \etal (1992) presented results of the search for variable stars on
photographic plates covering ${1\arcd\times1\zdot\arcd3}$ in the
north-eastern  part of the SMC. They found 17 RRab type stars and
estimated completeness of  the search to be 68\% (25 stars expected on
the entire field). The Smith \etal (1992) field overlaps slightly with
our eastern fields where we detected 59 RRab  stars. Simple
recalculation of the area of Smith \etal (1992) field and OGLE-II 
eastern, SMC$\_$E, fields (0.44 square degree) leads to 174 RRab-type
objects  expected in the 1.3 square degree region of similar stellar
density as OGLE-II  eastern fields. The Smith \etal (1992) field,
located farther to the North than OGLE-II fields,  has certainly much
lower stellar density -- by a factor of 3--5 as indicated from  our
tests performed on the Digitized Sky Survey images. Therefore we may 
conclude that our sample of the SMC RR~Lyr stars is reasonable complete,
and Smith  \etal (1992) estimate of completeness is likely
underestimated. For the LMC  fields, in which RRab stars are on average
0.4~mag brighter, the completeness of our  samples of RR~Lyr stars
should be even better. 

Completeness of the Carina dwarf galaxy RR~Lyr sample is probably a
little worse. We checked position of our 15 stars on the color-magnitude
diagram (CMD) of the Carina dwarf galaxy (Fig.~5) and found about 50\%
more stars in the region of horizontal branch which is occupied by our
RR~Lyr stars. However, the mean magnitudes of these extra stars are
similar to those of stars from our sample and we are convinced that our
present sample is representative enough for the Carina galaxy. The mean
{\it V} and {\it I}-band magnitudes of the Carina dwarf galaxy RR~Lyr
stars will be refined when the final RR~Lyr search in our field is
completed but it is unlikely that present figures will change by more
than 0.02~mag.

\Subsection{Relative Distance Scale
$d_{GB}^{RR}:d_{LMC}^{RR}:d_{SMC}^{RR}$ from RR~Lyr Stars}

Before we proceed to determination of the distance scale with the RR~Lyr
stars an important problem of constancy of the mean absolute magnitudes
of RR~Lyr stars in different stellar populations should be considered.
It is generally accepted that the {\it V}-band absolute magnitude of
RR~Lyr stars is a linear function of metallicity with the slope equal to
0.15--0.20~mag/dex. (Carney, Storm and Jones 1992, Skillen \etal 1993).
The most controversial point of this relation is its zero point. For the
rest of the paper we assume the following relation between the mean
absolute  {\it V}-band  magnitude of RR~Lyr stars and metallicity:

$$M_V^{\rm RR}={(0.18\pm0.03)}\times {\rm [Fe/H]}^{\rm RR} + {\rm const}
\hfill \eqno{(1)}$$

It should be noted that such a relation was derived  based on
observations of RR~Lyr stars from the Galaxy. We assume, as it is
generally accepted, that it holds also in other galaxies like the
Magellanic Clouds or the Carina dwarf galaxy. This is always a source of
some uncertainty but the problem concerns all standard candles which are
usually calibrated in our Galaxy and it is assumed that they share the
same properties in other objects.

What is the mean metallicity of our samples? In general all our samples 
contain the field RR~Lyr stars which have larger metallicities than
those  found in globular clusters. For the Galactic bulge sample the
situation is clear  -- Walker and Terndrup (1991) carried out a high
quality spectroscopic survey  and determined the mean metallicity of the
Baade's Window RR~Lyr stars (59 star sample) equal to ${\rm [Fe/H]^{\rm
RR}=-1.0}$ with small dispersion: 0.16~dex.  Similar result from Fourier
analysis of the OGLE-I RR~Lyr stars light curves,  ${\rm [Fe/H]^{\rm
RR}=-1.14\pm0.24}$, was obtained by Morgan, Simet and Bargenquast
(1998).

Little, however, is known about metallicities of field RR~Lyr stars in
the  Magellanic Clouds (\cf Olszewski, Suntzeff and Mateo 1996). Good
quality,  spectroscopic surveys of these stars have yet to be done,
which now,  with big telescopes and huge  samples of these objects
emerging as by-products  of the microlensing searches, can easily be
obtained.

In general it seems that the field RR~Lyr stars in the Magellanic Clouds
are more metal rich than those from globular clusters, similarly as in
the Galaxy.  Previous metallicity estimates range from ${-1.3}$~dex to
${-1.8}$~dex (Alcock  \etal 1996, Hazen and Nemec 1992) for different
fields in the LMC and from ${-1.6}$~dex to ${-1.8}$~dex for the SMC
fields (Smith \etal 1992, Butler \etal 1982). Therefore we adopted
metallicity equal to ${-1.6\pm0.2}$~dex for the LMC and
${-1.7\pm0.2}$~dex for the SMC RR~Lyr stars.

The mean period of the RR~Lyr stars is used often as an indicator of
the  metallicity of the sample. The mean periods of our samples are
0.549, 0.571  and 0.581 day for the Galactic bulge, LMC and SMC samples,
respectively. Although  deviations from the relation "mean period" --
metallicity can be  large in individual cases, these figures are
consistent with metallicities adopted in our study. We stress again that
a good  metallicity survey of Magellanic Clouds RR~Lyr stars would be of
great importance to clarify the problem.

The metallicity of RR~Lyr stars in Carina dwarf galaxy is estimated to
be ${-2.2}$~dex (Smecker-Hane \etal 1994).

\MakeTable{lcccccc}{12.5cm}{Mean {\it V}-band brightness of RR~Lyr stars
reduced to the Galactic bulge metallicity}
{
          & $V_0^{\rm RR}$ & $\sigma_{V_0}^{\rm ST}$ & 
          $\sigma_{V_0}^{\rm SYS}$ & $V_0^{\rm RR@GB}$ & 
          $\sigma_{\rm RR@GB}^{\rm SYS}$ & ${\rm [Fe/H]^{RR}}$\\
&&&&&&\\
GB        & 15.41 & 0.03  & 0.06 & 15.41 & 0.07 & $-1.0\pm0.1$ \\
LMC       & 18.86 & 0.04  & 0.08 & 18.97 & 0.09 & $-1.6\pm0.2$ \\
SMC       & 19.41 & 0.03  & 0.08 & 19.54 & 0.09 & $-1.7\pm0.2$ \\
Carina DG & 20.49 & 0.04  & 0.06 & 20.71 & 0.08 & $-2.2\pm0.2$ \\
}

Combining the metallicity differences with the absolute
magnitude-meta\-llicity relation we find that, on average, the LMC, SMC
and Carina dwarf galaxy  samples should  be about ${0.11\pm0.04}$~mag,
${0.13\pm0.04}$~mag and $0.22\pm0.05$~mag  brighter  than the Galactic
bulge one, respectively (Eq.~1). In Table~6 we present the mean {\it
V}-band magnitudes of the LMC, SMC and Carina RR~Lyr samples reduced to
the Galactic bulge metallicity, $V_0^{\rm  RR@GB}$, by applying above
corrections.

The possible systematic errors, $\sigma_{V_0}^{\rm SYS}$,  of the mean
magnitudes of our RR~Lyr samples  come mostly  from extinction
uncertainties. In the case of the Galactic bulge  the Stanek (1996) map
zero point was corrected with two independent methods  (Gould, Popowski
and Terndrup 1997, and Alcock \etal 1998b) yielding exactly  the same
corrections, thus indicating that the average systematic {\it V}-band
extinction  error of our sample should be smaller than 0.06~mag. With
uncertainties of the  zero point of the photometry the total systematic
uncertainty is 0.07~mag. In  the case of the Magellanic Clouds
extinction error should not exceed 0.08~mag. Combining this with
uncertainty of metallicity  difference between our Magellanic Cloud
samples and the Galactic bulge sample we obtain  the total systematic
error, $\sigma_{RR@GB}^{\rm SYS}$, to be 0.09~mag for both Magellanic
Clouds. For the Carina dwarf galaxy the {\it V}-band extinction
uncertainty is 0.06~mag, and $\sigma_{RR@GB}^{\rm SYS}$ 0.08~mag.

Now, we can proceed to determination of the relative distance scale
resulting from RR~Lyr photometry. Taking  the LMC distance as an unit,
we find the following relative distance scale
$d_{GB}^{RR}:d_{LMC}^{RR}:d_{SMC}^{RR}$

$$(0.194\pm0.010) : 1.00 : (1.30\pm0.08)$$

\Subsection{Absolute Calibration of the Relative RR~Lyrae Distance Scale} 

The absolute calibration of RR~Lyr stars has been a subject of many
disputes and controversies for many years. The RR~Lyr stars are too far
to have  reliable direct parallaxes even from the Hipparcos satellite.
At present the most  reliable calibration seems to come from statistical
parallaxes derived from  measurements of about hundred and fifty stars
with both ground based and  Hipparcos proper motions  (series of papers
by Gould and Popowski 1998 and references therein). After very careful
analysis of possible systematic errors of their large RR~Lyr stars
sample, Gould and Popowski (1998)  derive the mean absolute magnitude of
the RR~Lyr stars to be equal  ${M_V^{\rm RR}=0.77\pm0.13}$ at
metallicity ${\rm [Fe/H]^{\rm RR}=-1.6}$~dex. Similar  results but from
smaller sample were also obtained by Fernley \etal (1998). 

We may attempt to calibrate our RR~Lyr distance scale using Gould and
Popowski (1998) calibration and our Galactic bulge RR~Lyr sample for
which we have  precise metallicity determination from Walker and
Terndrup (1991) and to which metallicity the mean {\it V}-band
magnitudes of RR~Lyr stars in our objects were reduced (Table~6).

For the  metallicity of the Baade's Window RR~Lyr variables equal to
${-1.0}$~dex and  the mean slope of the $M_V^{\rm RR}$ -- metallicity
relation for RR~Lyr stars equal to 0.18 mag/dex  (Eq.~1) we obtain
${M_V^{\rm RR}=0.88\pm0.14}$~mag. Distance moduli are then derived by
subtracting that value from the mean {\it V}-band brightness reduced to
the bulge metallicity (column~4 Table~6) and are listed in Table~7. The
main component in the total error budget, $\sigma^{\rm TOT}$, comes 
from uncertainty of the Gould and Popowski (1998) calibration with much 
smaller contribution from statistical error of the mean {\it V}-band
magnitude of our RR~Lyr  samples and its systematic uncertainty.

\MakeTable{lcccc}{12.5cm}{RR~Lyr distances}
{
            & ${\rm [Fe/H]^{RR}}$ & $m-M$ & $\sigma^{\rm TOT}$ & $d$\\ 
            &                     &       &       & [kpc]\\       
&&&&\\
GB          &  $-1.0$             & 14.53 & 0.15  & ~8.1 \\
LMC         &  $-1.6$             & 18.09 & 0.16  & 41.5 \\
SMC         &  $-1.7$             & 18.66 & 0.16  & 54.0 \\
Carina DG   &  $-2.2$             & 19.83 & 0.15  & 92.5 \\
}

\Section{Red Clump Distance Scale: Galactic Bulge -- LMC -- SMC} 

Recent determinations of distances to the Galactic bulge (Paczy{\'n}ski
and  Stanek 1998), LMC (Udalski \etal 1998, Stanek, Zaritsky and Harris
1998) and SMC (Udalski  \etal 1998) with a newly developed red clump
stars method (Paczy{\'n}ski and  Stanek 1998) enable us to compare the
distance scale to all three objects  obtained with the RR~Lyr stars in
the previous Section with the red clump distances. Our selection of the
same lines-of-sight for the RR~Lyr and the red clump  determinations
makes both similarly affected by possible  systematic errors resulting
from interstellar extinction making such a comparison more reliable. The
statistical errors are almost an order of magnitude  smaller for the red
clump stars which are much more numerous than RR~Lyr stars  but even for
the latter they are of the order of only 0.03 mag, much smaller  than
systematic errors. 

\MakeTable{lccccl}{12.5cm}{Red clump mean luminosity}
{
            & $I_0^{\rm RC}$ & $\sigma_{I_0}^{\rm ST}$ & 
           $\sigma_{I_0}^{\rm SYS}$ & ${\rm [Fe/H]^{RC}}$ & References\\
&&&&&\\
GB          & 14.32   & 0.009 & 0.04 &  $+0.2\pm0.2$ & Paczy{\'n}ski and Stanek (1998)\\
LMC         & 17.85   & 0.004 & 0.05 &  $-0.6\pm0.2$ & Udalski \etal (1998) \\
SMC         & 18.33   & 0.003 & 0.05 &  $-0.8\pm0.2$ & Udalski \etal (1998) \\
M31 Halo    & 24.24   & 0.010 & 0.05 &  $-0.5\pm0.2$ & Stanek and Garnavich (1998) \\
M31 Cluster & 24.22   & 0.013 & 0.05 &  $-0.7\pm0.2$ & Stanek and Garnavich (1998) \\
Carina DG   & 19.44   & 0.005 & 0.04 &  $-1.9\pm0.2$ & this paper \\
}

In Table~8 we list the mean {\it I}-band, extinction free magnitudes of
the red clump stars, $I_0^{\rm RC}$, for objects to which the red clump
method has already been  applied: the Galactic bulge, LMC, SMC and M31
(Paczy{\'n}ski and Stanek 1998,  Udalski \etal 1998, Stanek and
Garnavich 1998) with their statistical, $\sigma_{I_0}^{\rm ST}$, and
systematic, $\sigma_{I_0}^{\rm SYS}$, errors. 

We added to our list of analyzed objects the Carina dwarf galaxy which
contains both old and intermediate age populations of stars similar to
the Magellanic Clouds. The Carina galaxy is also very metal poor which
makes it ideal for testing stability of the red clump luminosity in
different stellar environments.

In Table~8 we provide the mean {\it I}-band magnitude of the red clump
of the Carina dwarf galaxy determined in similar way as for the LMC and
SMC (Udalski \etal 1998). Fig.~5 presents the {\it I} \vs $(V-I)$ CMD of
the Carina galaxy based on our observations. Fig.~6 shows the luminosity
function histogram in 0.04~mag bins of stars in the Carina galaxy  red
clump region and the Gaussian function with second order polynomial
background fitted to the data (Udalski \etal 1998). We assumed the {\it
I}-band extinction equal to $0.12\pm0.03$ (Mighell 1997) and we selected
stars from the range: $0.7 < (V-I)_0 < 0.9$~mag and $20.5 < I_0 < 18.5$,
in total 696 stars. The color range  is somewhat different than in
previous determinations of the red clump mean luminosity, but because of
low metallicity and compactness of the red clump in the Carina galaxy
its large part falls blueward of the blue limit of previous
determinations. For comparison we also determined the mean magnitude of
the red part of the red clump falling in the range $0.8 < (V-I)_0 <
0.9$~mag. The mean magnitude of such a sample was 19.42~mag \ie 0.02~mag
brighter than the mean magnitude of the entire clump.

\Subsection{{\bb I}-band Mean Luminosity of the Red Clump Stars} 

The most controversial point of the red clump stars method of distance 
determination is the assumption of the same mean luminosity of the red
clump  stars in populations of different age, metallicity etc. Empirical
evidences  like constant {\it I}-band magnitude of the red clump stars
over the wide range of  observed colors (Paczy{\'n}ski and Stanek 1998),
similar distance  determination for different population fields in M31
(Stanek and Garnavich  1998) seem to suggest that this is indeed the
case. Also some models of the  red clump stars confirm weak dependence
of the red clump luminosity on age  (Castellani, Chieffi and Straniero
1992). On the other hand models of Bertelli  \etal (1994) predict much
stronger dependence of the red clump luminosity  on metallicity and age
reaching up to a few tens of magnitude for the  populations as different
as our local environment and Magellanic Cloud stars.  Cole (1998)
suggested that population effects can account even for 0.6~mag for 
different stellar populations. Girardi \etal (1998) obtained about 0.24
mag brighter red clump in the LMC than for stars from solar
neighborhood. Thus it is very important to verify empirically how well
the assumption of the constant mean luminosity of  the red clump holds. 

Having two potential standard candles in our four different objects,
Galactic bulge, LMC, SMC and Carina dwarf galaxy, we may compare them
and assuming that the RR~Lyr are, as commonly accepted, good standard
candles check how stable is the mean {\it I}-band magnitude of the red
clump stars in different populations. It should be stressed here, that
although both RR~Lyr stars  and the red clump giants have similar
internal structure -- both are He-core, H-shell burning stars -- we
compare two completely different stellar populations.  The field RR~Lyr
stars are the old objects with small metallicities. On the other hand
the red clump consists from much younger stars  but  with different
formation histories and metallicities in all our four objects.  The mean
metallicity of our red clump samples covers much wider range from the 
metal richest Galactic bulge with metallicity ${(-0.3\div0.3)}$~dex (\cf
Table~6 Minitti \etal 1995), through the LMC ${(-0.8\div-0.5)}$~dex
(Bica \etal 1998) and  SMC ${(-1.0\div-0.7)}$~dex (Olszewski, Suntzeff
and Mateo 1996) to Carina dwarf galaxy ${-1.9}$~dex (Mighell 1997). Also
other evidences like different spatial distribution  of both kinds of
stars in the Galactic bulge strongly suggest completely  different
populations -- the red clump stars are grouped in the bar (Stanek  \etal
1997), while the RR~Lyr stars belong to the spherical halo objects
(Alcock  \etal 1998a).

If we assume that the mean {\it V}-band magnitude of  the RR~Lyr stars
reduced to the same -- Galactic bulge -- metallicity is  constant, we
might treat the  mean reduced magnitudes of the RR~Lyr stars (column 4
of Table~6) as a reference point. This assumption is  supported by many
evidences like \eg similar luminosity of globular cluster and field
RR~Lyr stars (Catelan 1998). 

\MakeTable{lccccc}{12.5cm}{Red Clump minus RR~Lyr luminosity}
{
          & $I_0^{\rm RC}-V_0^{\rm RR@GB}$ & $\sigma_{\rm RC-RR@GB}^{TOT}$
          & ${\rm [Fe/H]^{RC}}$\\
&&&&&\\
GB        &  -1.09  & 0.04     & $+0.2\pm0.2$\\
LMC       &  -1.12  & 0.06     & $-0.6\pm0.2$\\
SMC       &  -1.21  & 0.06     & $-0.8\pm0.2$\\
Carina DG &  -1.27  & 0.07     & $-1.9\pm0.2$\\
}

Differences between the mean {\it I}-band brightness of the red clump
stars and reduced to the bulge metallicity, mean {\it V}-band luminosity
of the RR~Lyr stars, ${I_0^{\rm RC}-V_0^{\rm RR@GB}}$, with their total
errors, $\sigma_{\rm RC-RR@GB}^{TOT}$ are listed in Table~9. The total error
includes statistical error, systematic --  $E(V-I)$ reddening -- error
and uncertainty of the RR~Lyr brightness-metallicity relation (Eq.~1). 

Fig.~7 presents the differences as a function of metallicity of the red
clump stars, ${\rm [Fe/H]}^{\rm RC}$. There is a clear trend of larger 
difference for metal poorer objects suggesting somewhat brighter red
clump in metal deficient objects.  If our reference point defined by the
RR~Lyr mean {\it V}-band  magnitude at the Galactic bulge metallicity is
correct we observe an apparent  correlation between the mean red clump
magnitude and metallicity of the  object. We can therefore attempt to
calibrate the red clump mean magnitude and  the best, linear fit to the
data is:  

$$M^{\rm RC}_I=(0.09\pm0.03)\times{\rm [Fe/H]}^{\rm RC}+{\rm const}\,.
\eqno{(2)}$$

We are only interested in the slope of this relation, as we will
calibrate it very precisely in the next Subsection. It should be noted
that despite the fact that our calibration is based on four points only,
each of these points is very  well determined -- based on thousands of
red clump stars and tens of RR~Lyr  stars. They also cover a very wide
range of metallicities.  We hope that in the future more points can be
added to this calibration when for instance RR~Lyr stars are detected in
the Leo~I dwarf galaxy (Lee \etal 1993) or similar objects. 

The slope of the {\it I}-band brightness-metallicity relation for the
red clump stars we derived is much smaller than analogous slope of the
mean {\it V}-band brightness-metallicity relation for RR~Lyr stars
(Eq.~1) indicating much weaker dependence of red clump luminosity on
metallicity.

The dependence of the red clump {\it I}-band luminosity on the
metallicity was expected  from some  evolutionary models of the red
clump  stars (\eg Bertelli \etal  1994, Cole 1998).  For example, Cole
(1998) obtained  ${0.21\pm0.07}$~mag/dex slope of the
brightness-metallicity  relation for the red clump stars. Girardi \etal
(1998) predicted the difference of 0.24~mag between the absolute
luminosity of the red clump in the LMC and the local Hipparcos sample.
Our difference of the red clump absolute magnitude between the LMC and
the local Hipparcos sample is only 0.05~mag and the slope of
brightness-metallicity  relation 0.09 mag/dex.  Thus, it seems that
theoretical predictions are by factor of at least two too large
comparing with our empirical determination. It should be, however, noted
that the slope of the red clump stars relation is tied with the slope of
the RR~Lyr relation in the sense that larger slope of RR~Lyr relation
means larger slope for the red clump relation and {\it vice versa}. If,
for instance, we assume that the slope of the RR~Lyr
brightness-metallicity relation (Eq.~1) is as large as 0.37 mag/dex, as
sometimes suggested (Feast, 1997), the slope of the red clump relation
would increase to 0.20 mag/dex resulting in somewhat longer distances to
our objects than derived in the next Subsection (distance moduli larger
by $\approx 0.08$~mag for the Magellanic Clouds and $\approx 0.2$~mag
for the Carina galaxy).

\Subsection{Absolute Calibration of the Red Clump Luminosity} In the
case of the red clump stars the absolute calibration can be obtained 
with very good precision and practically the red clump stars are the
only  standard candle calibrated {\it via} a direct measurement.
Paczy{\'n}ski and  Stanek (1998) proposed a local red clump giants with
precise parallaxes  (accuracy better than 10\%) measured by Hipparcos as
a calibrator of the red  clump distances. The large number of such
stars, precise distances and  photometry make them ideal for that
purpose. The best calibration from the  distance (${d<70}$~pc) limited
sample (${>200}$ objects) is ${M_I^{\rm  RC}=-0.23\pm0.03}$~mag
(Stanek and Garnavich 1998). Assuming that the local  sample mean
metallicity is equal to solar, ${\rm [Fe/H]}^{\rm RC}=0.0$, we may use
the  Hipparcos data to determine the zero point of our
brightness-metallicity  relation for the red clump stars: 

$$M^{\rm RC}_I=(0.09\pm0.03)\times{\rm [Fe/H]}^{\rm RC}-0.23\pm0.03\,.
\eqno{(3)}$$

\MakeTable{lccccc}{12.5cm}{Red clump distances}
{
            & ${\rm [Fe/H]^{RC}}$ & Old   & Revised & $\sigma^{\rm TOT}$ & $d$ \\
            &                     & $m-M$ & $m-M$   &                   & [kpc]\\
&&&&&\\
GB          & $+0.2$              & 14.55 & 14.53   & 0.06  &  8.1 \\
LMC         & $-0.6$              & 18.08 & 18.13   & 0.07  & 42.3 \\
SMC         & $-0.8$              & 18.56 & 18.63   & 0.07  & 53.2 \\
M31 halo    & $-0.5$              & 24.47 & 24.52   & 0.07  & 802  \\
M31 cluster & $-0.7$              & 24.45 & 24.51   & 0.07  & 798  \\
Carina DG   & $-1.9$              &   --  & 19.84   & 0.07  & 92.9 \\
}

In Table~10 we list the new, revised distance moduli to objects to which
they  were determined before with the assumption of constant {\it
I}-band  magnitude of the red clump. Old distances are also included.
$\sigma^{\rm TOT}$ corresponds to the total error of the distance
determination: statistical, calibration and systematic errors. Table~10
also contains distance modulus to the  metal poor Carina dwarf galaxy
for  comparison of the red clump method distance determination with
other methods. 

As can be seen from Table~10, the differences between the old and new
distances  are small, at most of the order of 0.05--0.07~mag because of
small overall  metallicity range of objects analyzed so far and small
slope of the  brightness-metallicity relation for the red clump stars.
Much larger  discrepancy, reaching ${\approx0.2}$~mag would be expected
for the metal poor  objects. It should be noted here that the new
absolute distances, although calibrated independently of the RR~Lyr
distances, are somewhat dependent on RR~Lyr stars.  The
brightness-metallicity relation for RR~Lyr was assumed to bring the
RR~Lyr luminosities to the reference Galactic bulge metallicity
brightness, for determination of the brightness-metallicity relation of
the red clump stars.

The distance modulus to the most important astrophysical object, the 
LMC, can be now determined very precisely, as the calibration error is
small  -- about 0.04~mag, and systematic error can be reduced to
0.05~mag  based on similar results of Udalski \etal (1998) and Stanek,
Zaritsky and  Harris (1998). Thus, our final distance modulus of the LMC
is $m-M=18.13\pm0.07$~mag. Revised distance moduli to the SMC and
Galactic bulge are  $m-M=18.63\pm0.07$~mag and
$m-M=14.53\pm0.06$~mag, respectively. 

Finally, we can compare the distance to the metal poor Carina dwarf
galaxy determined with the revised red clump method with other results.
The most recent determination of the distance to the Carina galaxy was
obtained by  Mighell (1997) who determined the distance modulus equal to
${m-M=19.87\pm0.11}$~mag. This determination is in excellent agreement
with both our results -- from the red clump  and RR~Lyr stars.

\Section{Discussion}

We analyzed and determined the distance scale to three objects, the
Galactic  bulge, LMC and SMC, with two different standard candles: the
red clump stars  and RR~Lyr variables, based on the huge databases of
photometric measurements  from the OGLE-II microlensing search. From the
photometry of RR~Lyr stars in  the Galactic bulge, LMC and SMC we obtain
the following relative distance scale
$d_{GB}^{RR}:d_{LMC}^{RR}:d_{SMC}^{RR}$

$$(0.194\pm0.010) : 1.00 : (1.30\pm0.08)\,.$$

This scale is based on direct photometric measurements with an
appropriate  metallicity corrections applied to each sample of stars and
is independent of absolute calibration. 

The first test which we can immediately perform is the verification of
the  "long" distance scale to the LMC, ${m-M=18.50}$, \ie 50.1~kpc. If
we assume  that distance to the LMC, then the distance to the Galactic
bulge must be  ${9.72\pm0.50}$~kpc. This value seems to be in
contradiction with practically  all present determinations of the
distance to the Galactic center,  ${8.0\pm0.5}$~kpc (Reid 1993). On the
other hand even that crude calibration  -- the mean distance to the
Galactic center -- suggests "short" distance scale  to the LMC --
about 41.2~kpc (${m-M=18.08}$~mag).

Measurements of two standard candles in the same lines-of-sight: to the
Galactic bulge, LMC, SMC and Carina dwarf galaxy enable us to compare
them and -- adopting the  RR~Lyr luminosity as a reference point -- to
analyze the mean  luminosity of the red clump stars in objects with
different ages, star  formation histories and metallicities. Moreover,
our reference point, RR~Lyr  stars, belong to completely different, much
older, population   with  different metal content and kinematic
properties. This fact combined with  good standard candle properties of
RR~Lyr stars -- small dispersion of the mean absolute luminosity as
observed in globular clusters, similar properties  of RR~Lyr stars in
different environments (cluster and field objects) and well constrained
slope of the brightness-metallicity  dependence  -- make RR~Lyr stars a
very good reference for the red clump mean brightness comparison. 

Our analysis shows that the assumption of the constant red clump
luminosity  holds with accuracy of about 0.09~mag for populations
differing with mean  metallicity by up to 1~dex. We derive a calibration
of the red  clump mean {\it I}-band magnitude \vs metallicity based on
analysis of four objects, the  Galactic bulge, LMC, SMC and Carina dwarf
galaxy, covering large range of  metallicities: ${+0.2\div-1.9}$~dex.
The slope of this calibration seems to be well constrained,  the zero
point can be determined very precisely based on the local, Hipparcos 
measured red clump stars. The calibration is thus much more reliable
than that  of any other standard candle. The red clump stars method can
be now safely applied  even to very metal poor objects. 

Our empirical calibration of the red clump mean {\it I}-band magnitude
shows much smaller dependence on metallicity than resulting from
theoretical models (Cole 1998, Girardi \etal 1998). This discrepancy can
be a result of many difficult to verify assumptions (\eg star formation
history in the Magellanic Clouds, helium content) used for modeling of
the red clump.

We also do not see any significant dependence of the red clump
luminosity on age as residuals from our  brightness-metallicity relation
are negligible.  Models predict much brighter red clump  for younger
stars, younger than ${\approx2}$~Gyr, (Girardi and Bertelli 1998)  which
is indeed observed in the LMC young clusters. Brighter, young red clump
stars are also found as the vertical extension of the red clump in the
field color-magnitude diagrams (Zaritsky and  Lin 1997, Beaulieu and
Sackett 1998, Udalski \etal 1998). For older stars  luminosity is
practically age independent which indicates that the vast majority of
stars in the red clump of  the analyzed objects, Galactic bulge, LMC,
SMC and Carina galaxy must be older than 2~Gyr. Empirical tests of the
red clump brightness-age relation will be presented and discussed in the
next paper of the series.   

In Table~10 we present revised distance moduli to the Magellanic Clouds
and  Galactic bulge obtained with the modified red clump method. The
differences  between the old and revised distance moduli are small, not
exceeding 0.07~mag. The revised red clump method confirms the previous
result of the "short" distance scale to the LMC (Udalski \etal 1998,
Stanek, Zaritsky and Harris  1998). 

The absolute distances to the Galactic bulge, LMC, SMC and Carina galaxy
can also be  independently determined with the RR~Lyr relative distance
scale and the absolute  magnitude calibration for RR~Lyr stars. Such a
calibration is less accurate than calibration of the red clump method,
as even the closest galactic RR~Lyr  stars are too distant to be
measured directly. The most reliable calibration of the  RR~Lyr absolute
magnitude of Gould and Popowski (1998) gives the distance  moduli listed
in Table~7. They are in excellent agreement with determination from the
red clump stars method. 

Although our calibration of the brightness-metallicity relation of the
red clump stars based on the RR~Lyr stars as the reference point makes
the relative distance scales determined from the red clump and RR~Lyr
stars fully dependent, we have to stress that the absolute calibrations
of both distance indicators are absolutely independent. The red clump
calibration is based on the nearby red clump stars measured by the
Hipparcos satellite, while the calibration of the RR~Lyr stars comes
from much older, metal deficient stars, belonging to other population
and located in completely different parts of the Galaxy. Also the
methods of calibration are completely independent: trigonometric
parallaxes for the red clump stars and statistical parallaxes for RR~Lyr
stars. Absolute distances determined by both, so different distance
indicators are, however, in excellent agreement which strongly confirms
that our approach is correct.

Very good agreement of the distance determination to the LMC with the 
red clump stars method and RR~Lyr stars, is a strong argument in favor
of the  "short" distance scale to the LMC. From other reliable distance
indicators only the  Cepheid P--L relation seems to favor the "long"
distance scale. As discussed by  Udalski \etal (1998) and Stanek,
Zaritsky and Harris (1998) we suspect that  the Cepheid P--L relation
requires significant revision. For instance, Sekiguchi and  Fukugita
(1998) suggested that metallicity effects may bring the Cepheid P--L
based distance to the LMC to "short" values.

We believe that our relative distance scale to the Galactic bulge, LMC
and SMC, as well as absolutely calibrated distances determined with two
distance indicators belonging to different populations of stars and
calibrated independently  are at present the most  accurate distance
determinations to these objects. However, we have to stress  here that
independent determinations with other precise techniques are still  of
great importance. Well detached eclipsing binary systems seem to be the
most  promising candidates for accurate testing the distance  scale
(Paczy{\'n}ski 1997). It is a bit distressing that at the end of 20th
century one of the most important   topics of the modern astrophysics,
determination of the distance to  the LMC -- the milestone for  the
extragalactic distance scale -- is a subject of controversy reaching as
much  as 15\%. 

{\bf Acknowledgements.} We would like to thank Prof. Bohdan Paczy\'nski
for  many encouraging and stimulating discussions and help at all stages
of the OGLE project. We thank Drs M.~Kubiak, M.~Szyma{\'n}ski and
K.Z.~Stanek for many remarks and comments. Part of observations of the
Carina galaxy analyzed in this paper was carried out by Dr.\
M.~Szyma{\'n}ski. The  paper was partly supported by the Polish KBN
grant 2P03D00814 to A.\ Udalski.  Partial support for the OGLE project
was provided with the NSF grant  AST-9530478 to B.~Paczy\'nski.

\newpage

\centerline{\bf Figure captions}

\vspace{1cm}

\noindent
Fig.~1. Sample light curves of the RR~Lyr stars from the Small
Magellanic Cloud (field -- SMC$\_$SC1). Ordinate is phase (maximum of
brightness corresponds to the phase 0.0) and abscissa the {\it I}-band
magnitude. For each star the number from the database and period in days
are also given.\\

\noindent
Fig.~2. Sample light curves of the RR~Lyr stars from the Large
Magellanic Cloud (field -- LMC$\_$SC14). Ordinate is phase (maximum of
brightness corresponds to the phase 0.0) and abscissa the {\it I}-band
magnitude. For each star the number from the database and period in days
are also given.\\

\noindent
Fig.~3. Distribution of the {\it I}-band mean, extinction free, 
magnitudes of the RR~Lyr stars in the Galactic bulge (Baade's Window). 
Bins are 0.07~mag wide.\\

\noindent
Fig.~4. Distribution of the {\it I}-band mean, extinction free, 
magnitudes of the RR~Lyr stars in the Magellanic Cloud fields. Bins are
0.07~mag wide. \\

\noindent
Fig.~5. $I-(V-I)$ Color-Magnitude Diagram of the Carina dwarf
galaxy. The star denotes the mean location of the RR~Lyr stars.\\

\noindent
Fig.~6. Luminosity function of the red clump stars in the Carina dwarf
galaxy. Bins are 0.04~mag wide. The solid line represents the best fit
of a Gaussian superimposed on the parabola function.\\

\noindent
Fig.~7. Difference between the mean {\it I}-band magnitude of the red
clump stars and the mean {\it V}-band luminosity of the RR~Lyr stars
reduced to metallicity of the Galactic bulge in the Galactic bulge (GB),
LMC, SMC and Carina dwarf galaxy plotted as a function of metallicity of
the red clump stars. The dotted line represents the best linear fit
given by Eq.~2. 


\newpage

\begin{references}
\refitem{Alcock  \etal}{1996}{\AJ}{111}{1146}
\refitem{Alcock  \etal}{1998a}{\ApJ}{492}{190}
\refitem{Alcock  \etal}{1998b}{\ApJ}{494}{396}
\refitem{Beaulieu, J.P., and Sackett, P.D.}{1998}{\AJ}{~}{in press,
(astro-ph/9710156)}
\refitem{Bertelli, G., Bressan, A., Chiosi, C., Fagotto, F., and Nasi., E.}
{1994}{\AA ~Suppl. Ser.}{106}{275}
\refitem{Bica, E., Dottori, H., and Pastoriza, M.}{1987}{\AA}{156}{261}
\refitem{Bica, E., Geisler, D., Clari\'a, J.J., and Piatti,
A.E.}{1998}{\AJ}{~}{in press, (astro-ph/9803167)}
\refitem{Butler, D, Demarque, P., and Smith, H.A.}{1982}{\ApJ}{257}{592}
\refitem{Carney, B.W., Storm, J., and Jones, R.V.}{1992}{\ApJ}{386}{663}
\refitem{Castellani, V., Chieffi, A., and Straniero, O.}{1992}{\ApJS}{78}{517}
\refitem{Catelan M.}{1998}{\ApJL}{495}{L81}
\refitem{Cole, A.A.}{1998}{\ApJL}{~}{in press, (astro-ph/9804110)}
\refitem{Feast, M.W.}{1997}{\MNRAS}{284}{761}
\refitem{Fernley, J., \etal}{1998}{\AA}{330}{512}
\refitem{Girardi, L., and Bertelli, G.}{1998}{\MNRAS}{~}
{in press, (astro-ph/9801145)}
\refitem{Girardi, L., Groenewegen, M.A.T., Weiss, A., and Salaris,
M.}{1998}{\MNRAS}{~} {submitted, (astro-ph/9805127)}
\refitem{Gould, A., and Uza, O.}{1998}{\ApJL}{494}{L118}
\refitem{Gould, A., Popowski, P., and Terndrup, D.T.}{1998}{ApJ}{492}{778}
\refitem{Gould, A., and Popowski, P.}{1998}{\ApJ}{~}{submitted, 
(astro-ph/9805176)} 
\refitem{Hazen, M.L, and Nemec, J.M.}{1992}{\AJ}{104}{111}
\refitem{Hurley-Keller, D., Mateo, M., and Nemec, J.}{1998}{\AJ}{~}{in
press, (astro-ph/9804058)}
\refitem{Landolt, A.U.}{1992}{\AJ}{104}{372}
\refitem{Laney, C.D, and Stobie, R.S.}{1994}{\MNRAS}{266}{441}
\refitem{Layden, A.C., Hanson, R.B., Hawley, S.L., Klemola, A.R., and Hanley,
C.J.}{1996}{\AJ}{112}{2110}
\refitem{Lee M.G., Freedman, W., Mateo, M., Thompson, I., Roth, M.,
and Ruiz, M.T.}{1993}{\AJ}{106}{1420}
\refitem{Madore, B.F., and Freedman, W.L.}{1998}{\ApJL}{492}{110}
\refitem{Mighell, K.J.}{1997}{\AJ}{114}{1458}
\refitem{Minitti, D., Olszewski, E.W., Liebert, J., White, S.D.M., Hill,
J.M., and Irwin, M.J.}{1994}{\MNRAS}{277}{1293}
\refitem{Morgan, S.M., Simet, M., and Bargenquast,
S}{1998}{\Acta}{48}{in press}
\refitem{Olszewski, E.W., Suntzeff, N.B., and  Mateo, M.}{1996}
{\it ARA\&A}{34}{511}
\refitem{Paczy\'nski, B.}{1997}{~}{~}{in: "The Extragalactic Distance Scale 
STScI Symposium", Baltimore, Cambridge University Press, 273 
(astro-ph/9608094)}
\refitem{Paczy\'nski B., and Stanek, K.Z.}{1998}{\ApJL}{494}{L219}
\refitem{Panagia, N., Gilmozzi, R., Kirshner, R.P., Pun, C.S.J.,
and Sonneborn, G.}{1997}{\it BAAS}{191}{19.09}
\refitem{Reid, M.J.}{1993}{\it ARA\&A}{31}{345}
\refitem{Reid, N., and Freedman, W.L.}{1994}{\MNRAS}{267}{821}
\refitem{Saha, A., Monet, D.G., and Seitzer, P.}{1986}{\AJ}{92}{302}
\refitem{Sekiguchi, M., and Fukugita, M.}{1998}{Observatory}{~}{in
press, (astro-ph/9707229)}
\refitem{Skillen, I., Fernley, J.A., Stobie, R.S., and Jameson,
R.F.}{1993}{\MNRAS}{265}{301}
\refitem{Smecker-Hane, T.A., Stetson, P.B., Hesser, J.E. and Lehnert,
M.D.}{1994}{\AJ}{108}{507}
\refitem{Smith H.A, Silbermann, N.A., Baird, S.R., and Graham, J.A.} 
{1992}{\AJ}{104}{1430}, 
\refitem{Stanek, K.Z.}{1996}{\ApJL}{460}{L37}
\refitem{Stanek, K.Z., Udalski, A., Szyma\'nski, M., Ka{\l}u{\.z}ny,J., 
Kubiak, M., Mateo, M., and Krzemi\'nski, W.}{1997}{\ApJ}{477}{163}
\refitem{Stanek K.Z., and Garnavich, P.M.}{1998}{\ApJL}{~}{in press,
(astro-ph/9802121)}
\refitem{Stanek K.Z, Zaritsky, D., and Harris, J.}{1998}{\ApJL}{~}{in
press, (astro-ph/9803181)}
\refitem{Udalski, A., Kubiak, M., and Szyma\'nski, M.}{1997}{\Acta}{47}{319}
\refitem{Udalski A., Szyma\'nski, M., Ka{\l}u{\.z}ny,J., Kubiak, M., and
Mateo, M.}{1993}{\Acta}{43}{69}
\refitem{Udalski, A., Szyma\'nski, M., Ka{\l}u{\.z}ny,J., Kubiak, M.,
Mateo, M., and Krzemi\'nski, W.}{1994}{\Acta}{44}{317}
\refitem{Udalski, A., Szyma\'nski, M., Ka{\l}u{\.z}ny,J., Kubiak, M.,
Mateo, M., Krzemi\'nski, W.}{1995a}{\Acta}{45}{1}
\refitem{Udalski, A., Olech, A., Szyma\'nski, M., Ka{\l}u{\.z}ny,J., Kubiak, M.,
Mateo, M., Krzemi\'nski, W.}{1995b}{\Acta}{45}{433}
\refitem{Udalski, A., Szyma\'nski, M., Kubiak, M., Pietrzy\'nski, G.,
Wo\'zniak, P., and {\.Z}ebru\'n, K.}{1998}{\Acta}{48}{1, (astro-ph/9803035)}
\refitem{Walker, A.R., and Terndrup, D.M.}{1991}{\ApJ}{378}{119}
\refitem{Zaritsky, D., and Lin, D.B.C.}{1997}{\AJ}{114}{2545}
\end{references}
\end{document}